# Branching Law for Axons

**Dmitri B. Chklovskii and Armen Stepanyants**
Cold Spring Harbor Laboratory
1 Bungtown Rd.
Cold Spring Harbor, NY 11724
*mitya@cshl.edu   stepanya@cshl.edu*

## Abstract

What determines the caliber of axonal branches? We pursue the hypothesis that the axonal caliber has evolved to minimize signal propagation delays, while keeping arbor volume to a minimum. We show that for a general cost function the optimal diameters of mother $(d_0)$ and daughter $(d_1, d_2)$ branches at a bifurcation obey a branching law: $d_0^{\nu+2} = d_1^{\nu+2} + d_2^{\nu+2}$. The derivation relies on the fact that the conduction speed scales with the axon diameter to the power $\nu$ ($\nu = 1$ for myelinated axons and $\nu = 0.5$ for non-myelinated axons). We test the branching law on the available experimental data and find a reasonable agreement.

## 1 Introduction

Multi-cellular organisms have solved the problem of efficient transport of nutrients and communication between their body parts by evolving spectacular networks: trees, blood vessels, bronchs, and neuronal arbors. These networks consist of segments bifurcating into thinner and thinner branches. Understanding of branching in transport networks has been advanced through the application of the optimization theory ([1], [2] and references therein). Here we apply the optimization theory to explain the caliber of branching segments in communication networks, i.e. neuronal axons.

Axons in different organisms vary in caliber from 0.1μm (terminal segments in neocortex) to 1000μm (squid giant axon) [3]. What factors could be responsible for such variation in axon caliber? According to the experimental data [4] and cable theory [5], thicker axons conduct action potential faster, leading to shorter reaction times and, perhaps, quicker thinking. This increases evolutionary fitness or, equivalently, reduces costs associated with conduction delays. So, why not make all the axons infinitely thick? It is likely that thick axons are evolutionary costly because they require large amount of cytoplasm and occupy valuable space [6], [7]. Then, is there an optimal axon caliber, which minimizes the combined cost of conduction delays and volume?

In this paper we derive an expression for the optimal axon diameter, which minimizes the combined cost of conduction delay and volume. Although the relative cost of delay and volume is unknown, we use this expression to derive a law describing segment caliber of branching axons with no free parameters. We test this law on the published anatomical data and find a satisfactory agreement.

## 2  Derivation of the branching law

Although our theory holds for a rather general class of cost functions (see Methods), we start, for the sake of simplicity, by deriving the branching law in a special case of a linear cost function. Detrimental contribution to fitness, $\mathfrak{C}$, of an axonal segment of length, $L$, can be represented as the sum of two terms, one proportional to the conduction delay along the segment, $T$, and the other - to the segment volume, $V$:

$$\mathfrak{C} = \alpha T + \beta V . \tag{1}$$

Here, $\alpha$ and $\beta$ are unknown but constant coefficients which reflect the relative contribution to the fitness cost of the signal propagation delay and the axonal volume.

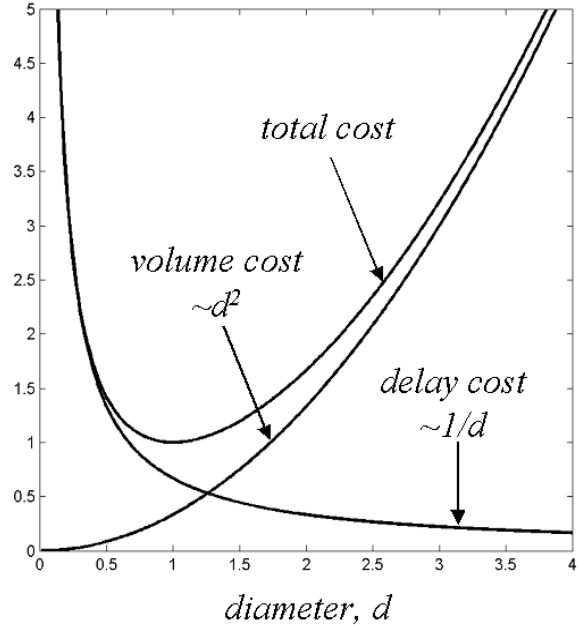

Figure 1: Fitness cost of a myelinated axonal segment as a function of its diameter. The lines show the volume cost, the delay cost, and the total cost. Notice that the total cost has a minimum. Diameter and cost values are normalized to their respective optimal values.

We look for the axon caliber $d$ that minimizes the cost function $\mathfrak{C}$. To do this, we rewrite $\mathfrak{C}$ as a function of $d$ by noticing the following relations: *i)* Volume,

$V = \frac{\pi}{4} L d^2$; ii) Time delay, $T = \frac{L}{s}$; iii) Conduction velocity $s = kd$ for myelinated axons (for non-myelinated axons, see Methods):

$$\mathfrak{C} = \alpha \frac{L}{s} + \beta L \frac{\pi}{4} d^2 = L\left(\frac{\alpha}{kd} + \frac{\beta \pi}{4} d^2\right). \tag{2}$$

This cost function contains two terms, which have opposite dependence on $d$, and has a minimum, Fig. 1.

Next, by setting $\frac{\partial \mathfrak{C}}{\partial d} = 0$ we find that the cost is minimized by the following axonal caliber:

$$d = \left(\frac{2\alpha}{\pi k \beta}\right)^{1/3}. \tag{3}$$

The utility of this result may seem rather limited because the relative cost of time delays vs. volume, $\alpha/\beta$, is unknown.

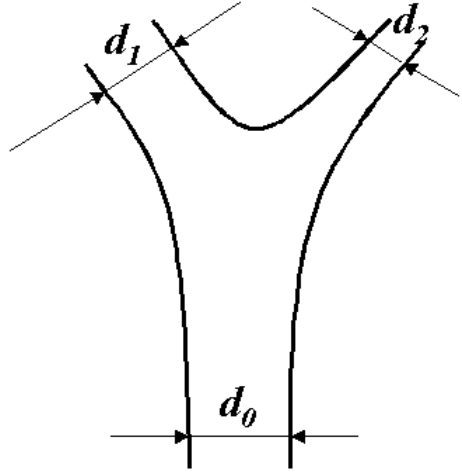

Figure 2: A simple axonal arbor with a single branch point and three axonal segments. Segment diameters are $d_0$, $d_1$, and $d_2$. Time delays along each segment are $t_0$, $t_1$, and $t_2$. The total time delay down the first branch is $T_1 = t_0 + t_1$, and the second - $T_2 = t_0 + t_2$.

However, we can apply this result to axonal branching and arrive at a testable prediction about the relationship among branch diameters without knowing the relative cost. To do this we write the cost function for a bifurcation consisting of three segments, Fig. 2:

$$\mathfrak{C} = \alpha_1(t_0 + t_1) + \alpha_2(t_0 + t_2) + \beta(V_0 + V_1 + V_2), \tag{4}$$

where $t_0$ is a conduction delay along segment 0, $t_1$ - conduction delay along segment 1, $t_2$ - conduction delay along segment 2. Coefficients $\alpha_1$ and $\alpha_2$ represent relative costs of conduction delays for synapses located on the two daughter branches and may be different. We group the terms corresponding to the same segment together:

$$\mathfrak{C} = [(\alpha_1 + \alpha_2)t_0 + \beta V_0] + [\alpha_1 t_1 + \beta V_1] + [\alpha_2 t_2 + \beta V_2]. \tag{5}$$

We look for segment diameters, which minimize this cost function. To do this we make the dependence on the diameters explicit and differentiate in respect to them. Because each term in Eq. (5) depends on the diameter of only one segment the variables separate and we arrive at expressions analogous to Eq.(3):

$$d_0 = \left(\frac{2(\alpha_1 + \alpha_2)}{k\beta\pi}\right)^{1/3}, \quad d_1 = \left(\frac{2\alpha_1}{k\beta\pi}\right)^{1/3}, \quad d_2 = \left(\frac{2\alpha_2}{k\beta\pi}\right)^{1/3}. \tag{6}$$

It is easy to see that these diameters satisfy the following branching law:

$$d_0^3 = d_1^3 + d_2^3. \tag{7}$$

Similar expression can be derived for non-myelinated axons (see Methods). In this case, the conduction velocity scales with the square root of segment diameter, resulting in a branching exponent of $2.5$.

We note that expressions analogous to Eq. (7) have been derived for blood vessels, tree branching and bronchs by balancing metabolic cost of pumping viscous fluid and volume cost [8], [9]. Application of viscous flow to dendrites has been discussed in [10]. However, it is hard to see how dendrites could be conduits to viscous fluid if their ends are sealed.

Rall [11] has derived a similar law for branching dendrites by postulating impedance matching:

$$d_0^{3/2} = d_1^{3/2} + d_2^{3/2}. \tag{8}$$

However, the main purpose of Rall's law was to simplify calculations of dendritic conduction rather than to explain the actual branch caliber measurements.

## 3  Comparison with experiment

We test our branching law, Eq.(7), by comparing it with the data obtained from myelinated motor fibers of the cat [12], Fig. 3. Data points represent 63 branch points for which all three axonal calibers were available. Eq.(7) predicts that the data should fall on the line described by:

$$\left(\frac{d_1}{d_0}\right)^\eta + \left(\frac{d_2}{d_0}\right)^\eta = 1, \tag{9}$$

where exponent $\eta = 3$. Despite the large spread in the data it is consistent with our predictions. In fact, the best fit exponent, $\eta = 2.57$, is closer to our prediction than to Rall's law, $\eta = 1.5$.

We also show the histogram of the exponents $\eta$ obtained for each of $63$ branch points from the same data set, Fig. 4. The average exponent, $\eta = 2.67$, is much closer to our predicted value for myelinated axons, $\eta = 3$, than to Rall's law, $\eta = 1.5$.

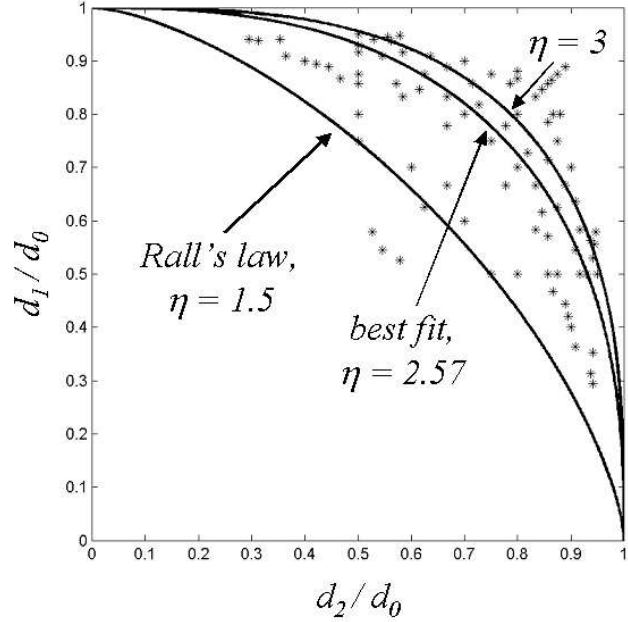

Figure 3: Comparison of the experimental data (asterisks) [12] with theoretical predictions. Each axonal bifurcation (with $d_1 \neq d_2$) is represented in the plot twice. The lines correspond to Eq.(9) with various values of the exponent: the Rall's law, $\eta = 1.5$, the best-fit exponent, $\eta = 2.57$, and our prediction for myelinated axons, $\eta = 3$.

Analysis of the experimental data reveals a large spread in the values of the exponent, $\eta$. This spread may arise from the biological variability in the axon diameters, other factors influencing axon diameters, or measurement errors due to the finite resolution of light microscopy. Although we cannot distinguish between these causes, we performed a simulation showing that a reasonable measurement error is sufficient to account for the spread.

First, based on the experimental data [12], we generate a set of diameters $d_0$, $d_1$ and $d_2$ at branch points, which satisfy Eq. (7). We do this by taking all diameter pairs at branch point from the experimental data and calculating the value of the third diameter according to Eq. (7). Next we simulate the experimental data by adding Gaussian noise to all branch diameters, and calculate the probability distribution for the exponent $\eta$ resulting from this procedure. The line in Fig. 4 shows that the spread in the histogram of branching exponent could be explained by Gaussian measurement error with standard deviation of $0.4 \mu m$. This value of standard deviation is consistent with $0.5 \mu m$ precision with which diameter measurements are reported in [12].

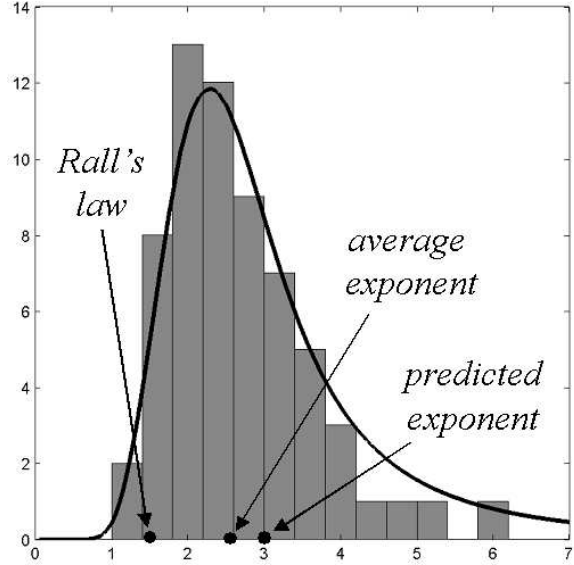

Figure 4: Experimentally observed spread in the branching exponent may arise from the measurement errors. The histogram shows the distribution of the exponent $\eta$, Eq. (9), calculated for each axonal bifurcation [12]. The average exponent is $\eta = 2.67$. The line shows the simulated distribution of the exponent obtained in the presence of measurement errors.

## 4 Conclusion

Starting with the hypotheses that axonal arbors had been optimized in the course of evolution for fast signal conduction while keeping arbor volume to a minimum we derived a branching law that relates segment diameters at a branch point. The derivation was done for the cost function of a general form, and relies only on the known scaling of signal propagation velocity with the axonal caliber. This law is consistent with the available experimental data on myelinated axons. The observed spread in the branching exponent may be accounted for by the measurement error. More experimental testing is clearly desirable.

We note that similar considerations could be applied to dendrites. There, similar to non-myelinated axons, time delay or attenuation of passively propagating signals scales as one over the square root of diameter. This leads to a branching law with exponent of $5/2$. However, the presence of reflections from branch points and active conductances is likely to complicate the picture.

## 5 Methods

The detrimental contribution of an axonal arbor to the evolutionary fitness can be quantified by the cost, $\mathfrak{C}$. We postulate that the cost function, $\mathfrak{C}$, is a monotonically increasing function of the total axonal volume per neuron, $V$, and all signal propagation delays, $T_j$, from soma to $j$-th synapse, where $j = 1, 2, 3, \ldots$:

$$\mathfrak{C}(V, T_1, T_2, T_3, \ldots) . \qquad (10)$$

Below we show that this rather general cost function (along with biophysical properties of axons) is minimized when axonal caliber satisfies the following branching law:

$$d_0^\eta = d_1^\eta + d_2^\eta \tag{11}$$

with branching exponent $\eta = 3$ for myelinated and $\eta = 2.5$ for non-myelinated axons.

Although we derive Eq. (11) for a single branch point, our theory can be trivially extended to more complex arbor topologies. We rewrite the cost function, $\mathfrak{C}$, in terms of volume contributions, $V_i$, of $i$-th axonal segment to the total volume of the axonal arbor, $V$, and signal propagation delay, $t_i$, occurred along $i$-th axonal segment. The cost function reduces to:

$$\mathfrak{C}(V_0 + V_1 + V_2, t_0 + t_1, t_0 + t_2). \tag{12}$$

Next, we express volume and signal propagation delay of each segment as a function of segment diameter. The volume of each cylindrical segment is given by:

$$V_i = \frac{\pi}{4} L_i d_i^2, \tag{13}$$

where $L_i$ and $d_i$ are segment length and diameter, correspondingly. Signal propagation delay, $t_i$, is given by the ratio of segment length, $L_i$, and signal speed, $s_i$. Signal speed along axonal segment, in turn, depends on its diameter as:

$$s_i = k d_i^\nu, \tag{14}$$

where $\nu = 1$ for myelinated [4] and $\nu = 0.5$ for non-myelinated fibers [5]. As a result propagation delay along segment $i$ is:

$$t_i = \frac{L_i}{k d_i^\nu}. \tag{15}$$

Substituting Eqs. (13), (15) into the cost function, Eq. (12), we find the dependence of the cost function on segment diameters,

$$\mathfrak{C}\left(\frac{\pi}{4} L_0 d_0^2 + \frac{\pi}{4} L_1 d_1^2 + \frac{\pi}{4} L_2 d_2^2, \frac{L_0}{k d_0^\nu} + \frac{L_1}{k d_1^\nu}, \frac{L_0}{k d_0^\nu} + \frac{L_2}{k d_2^\nu}\right). \tag{16}$$

To find the diameters of all segments, which minimize the cost function $\mathfrak{C}$, we calculate its partial derivatives with respect to all segment diameters and set them to zero:

$$\frac{\partial \mathfrak{C}}{\partial d_0} = \mathfrak{C}'_V \frac{\pi}{2} L_0 d_0 - \mathfrak{C}'_{T_1} \frac{\nu L_0}{k d_0^{\nu+1}} - \mathfrak{C}'_{T_2} \frac{\nu L_0}{k d_0^{\nu+1}} = 0$$

$$\frac{\partial \mathfrak{C}}{\partial d_1} = \mathfrak{C}'_V \frac{\pi}{2} L_1 d_1 - \mathfrak{C}'_{T_1} \frac{\nu L_1}{k d_1^{\nu+1}} = 0 \qquad (17)$$

$$\frac{\partial \mathfrak{C}}{\partial d_2} = \mathfrak{C}'_V \frac{\pi}{2} L_2 d_2 - \mathfrak{C}'_{T_2} \frac{\nu L_2}{k d_2^{\nu+1}} = 0$$

By solving these equations we find the optimal segment diameters:

$$d_0^{\nu+2} = \frac{2\nu\left(\mathfrak{C}'_{T_1} + \mathfrak{C}'_{T_2}\right)}{k\pi \mathfrak{C}'_V}, \quad d_1^{\nu+2} = \frac{2\nu \mathfrak{C}'_{T_1}}{k\pi \mathfrak{C}'_V}, \quad d_2^{\nu+2} = \frac{2\nu \mathfrak{C}'_{T_2}}{k\pi \mathfrak{C}'_V}. \qquad (18)$$

These equations imply that the cost function is minimized when the segment diameters at a branch point satisfy the following expression (independent of the particular form of the cost function, which enters Eq. (18) through the partial derivatives $\mathfrak{C}'_V$, $\mathfrak{C}'_{T_1}$, and $\mathfrak{C}'_{T_2}$):

$$d_0^\eta = d_1^\eta + d_2^\eta, \quad \eta = \nu + 2. \qquad (19)$$